\documentclass[useAMS,usenatbib,usegraphicx,letterpaper]{mn2e}
\usepackage{times,amssymb,hyperref,subfigure,epsfig,aas_macros}

\usepackage[totalwidth=480pt,totalheight=680pt,layoutvoffset=0.5cm]{geometry}

\begin{document}

\title[Star - Planet - Debris Disk Alignment]{Star - Planet - Debris Disk Alignment in
  the HD 82943 system: \\ Is planetary system coplanarity actually the norm?}


\author[G. M. Kennedy et al.]
{G. M. Kennedy\thanks{Email:
    \href{mailto:gkennedy@ast.cam.ac.uk}{gkennedy@ast.cam.ac.uk}}$^1$, M. C. Wyatt$^1$,
  G. Bryden$^2$, R. Wittenmyer$^3$, B. Sibthorpe$^4$ \\
  $^1$ Institute of Astronomy, University of Cambridge, Madingley Road, Cambridge CB3
  0HA, UK \\
  $^2$ Jet Propulsion Laboratory, California Institute of Technology, 4800 Oak Grove Drive, Pasadena, CA 91109, USA \\
  $^3$  Department of Astrophysics, School of Physics, University of NSW, 2052, Australia \\
  $^4$ SRON Netherlands Institute for Space Research, NL-9747 AD Groningen, The Netherlands \\
}
\maketitle

\begin{abstract}
  Recent results suggest that the two planets in the HD 82943 system are inclined to the
  sky plane by $20 \pm 4^\circ$. Here, we show that the debris disk in this system is
  inclined by $27 \pm 4^\circ$, thus adding strength to the derived planet inclinations
  and suggesting that the planets and debris disk are consistent with being aligned at a
  level similar to the Solar System. Further, the stellar equator is inferred to be
  inclined by $28 \pm 4^\circ$, suggesting that the entire star - planet - disk system is
  aligned, the first time such alignment has been tested for radial velocity discovered
  planets on $\sim$AU wide orbits. We show that the planet-disk alignment is primordial,
  and not the result of planetary secular perturbations to the disk inclination. In
  addition, we note three other systems with planets at $\gtrsim$10AU discovered by
  direct imaging that already have good evidence of alignment, and suggest that empirical
  evidence of system-wide star - planet - disk alignment is therefore emerging, with the
  exception of systems that host hot Jupiters. While this alignment needs to be tested in
  a larger number of systems, and is perhaps unsurprising, it is a reminder that the
  system should be considered as a whole when considering the orientation of planetary
  orbits.
\end{abstract}


\section{Introduction}\label{s:intro}

Planetary systems are known to emerge from the disk-like structures of gas and dust that
surround young stars. It has therefore generally been expected that, as in the Solar
System, all components of exo-planetary systems should share a common angular momentum
direction; the planets and debris disk should orbit in the same direction and in the same
plane as the stellar equator. Of course, the most well studied system, our Solar System,
is not perfectly aligned with a single plane. A variation of nearly $10^\circ$ when the
Sun's equator and Mercury's orbit are included suggests a benchmark for star - planet -
disk alignment in other systems.

The discovery of star - planet misalignment for transiting gas giants has been a
surprising counterpoint to the expectation of alignment. Though nearly all of the first
dozen transiting systems were found to be aligned \citep[see][and references
therein]{2009ApJ...696.1230F}, proof that alignment is not always the case
\citep[e.g.][]{2010A&A...524A..25T} has prompted theoretical work that attempts to
explain their existence
\citep[e.g.][]{2007ApJ...669.1298F,2011MNRAS.412.2790L,2012Natur.491..418B}. Misalignment
could be indicative of processes acting after the formation of the planetary system, and
be specific to the way in which some hot Jupiters form. For example, the planets could
originate on orbits that are aligned with the star, but be circularised after being
forced to low perihelia via long-term dynamical interactions with other planets or
stellar companions that excite their eccentricities and inclinations, naturally forming
misaligned systems \citep{2007ApJ...669.1298F}. Alternatively, the misalignment could
originate from a primordial misalignment of the gaseous protoplanetary disk
\citep{2011MNRAS.412.2790L,2012Natur.491..418B}, implying that hot Jupiters could have
migrated through the gas disk to their observed locations without experiencing strong
dynamical interactions with other bodies. Since the stellar rotation-planet orbit
alignment has only been tested outside the Solar System using the Rossiter-McLaughlin
effect and starspot occultation \citep{2011ApJ...740L..10N}, measurements that are
generally only possible on close in transiting planets, it is not yet possible to tell if
the observed misalignment is representative of planetary systems in general.

One prediction of the primordially misaligned disk scenarios is that debris disks,
presumed to have their origins within the gaseous protoplanetary disk, could be
misaligned with their parent stars. However, the test for star - disk alignment has until
recently been much harder. It involves comparing the inclination of the star inferred
from the stellar radius, projected rotation velocity, and rotation period, to that of the
resolved debris disk that orbits that star. This test is not usually possible because
debris disks are only detected around $\sim$15\% of Sun-like stars, and until the launch
of \emph{Herschel}\footnote{\emph{Herschel} is an ESA space observatory with science
  instruments provided by European-led Principal Investigator consortia and with
  important participation from NASA.} few of these were resolved. In addition, the
position angle of the stellar inclination is rarely measured \citep[but
see][]{2009A&A...498L..41L}. Therefore, star-disk alignment is shown in a statistical
sense rather than for individual systems. In the cases where this test has become
possible, the conclusion is that the stellar and disk inclinations are generally similar,
and hence that both share the same orbital plane as the primordial protoplanetary disk
\citep[][Greaves et al, in preparation]{2011MNRAS.413L..71W}.\footnote{It is also
  possible to test binary orbital plane - disk alignment if the binary orbit is well
  characterised, see \citet{2010ApJ...710..462A} and
  \citet{2012MNRAS.421.2264K,2012MNRAS.426.2115K}.}

The final alignment test, that of planet - disk alignment is in general the least common
due to the rarity of systems in which it is possible. Curiously however, all three
systems with directly imaged planets (around A-stars) allow this test. Despite
uncertainties about the orbit and nature of the planet around Fomalhaut
\citep{2008Sci...322.1345K,2011MNRAS.412.2137K,2012ApJ...747..116J}, the current
understanding has Fomalhaut b consistent with, though by no means guaranteed to be,
aligned with the spectacular debris ring (which is inclined by 66$^\circ$, Kalas et al
submitted). In addition, the position angle of the stellar rotation axis of $65 \pm
3^\circ$ \citep{2009A&A...498L..41L} is perpendicular to the debris disk major axis of
$156 \pm 0.3^\circ$ \citep{2005Natur.435.1067K}, suggesting that the stellar equator is
also aligned with the ring. In the HR 8799 system, the favoured orbits are near face-on
\citep{2008Sci...322.1348M,2010ApJ...710.1408F,2009ApJ...694L.148L}, as is the debris
disk \citep{2009ApJ...705..314S}. In fact, the favoured planetary inclination of
13-23$^\circ$ is very similar to the debris disk inclination derived from \emph{Herschel}
observations (Matthews et al, in preparation). Further, HR 8799 itself is also inferred
to be nearly pole-on, with an inclination of 13-30$^\circ$
\citep{2009A&A...503..247R}. Finally, the planet around $\beta$ Pictoris is consistent
with being aligned with the edge-on disk \citep{2010Sci...329...57L,2011ApJ...736L..33C},
but may be slightly misaligned ($\sim$5$^\circ$), if it is the origin of the disk warp
seen at $\sim$70AU \citep{1997MNRAS.292..896M,2011ApJ...743L..17D}. Therefore, in the
cases where it is possible to test star - planet - disk alignment, at radial scales well
beyond the realm of hot Jupiters, alignment at our benchmark level is the conclusion in
all three cases.

In summary, with the caveat that some hot Jupiters may be misaligned with their host
stars due to their formation mechanism, it appears that, as expected, empirical evidence
of star - planet - disk alignment as the norm in planetary systems is emerging. However,
with only four cases that argue for alignment, and the example that the first hot
Jupiters were found to be aligned, more examples are clearly needed to test the
primordially misaligned models. The planets in the three aligned systems discussed above
are all at $\gtrsim$10AU around A-type stars, so tests at scales between the realm of
direct imaging and transits (i.e.  $\sim$AU scales), and around Sun-like stars are
especially lacking.

Here, we focus on alignment in the HD 82943 system, whose planets orbit the Sun-like host
star at $\sim$AU distances. Recent results from \citet{2013arXiv1306.0687T} suggest that,
assuming that their orbits are coplanar, the two giant planets in this system are
inclined to the sky plane by $i=20 \pm 4^\circ$. We show that the debris disk as resolved
by \emph{Herschel} imaging is inclined by $27 \pm 4^\circ$, thereby adding strength to
the inferred planet inclinations, and arguing that the planets and disk are aligned. In
addition, we show that the inferred stellar inclination is $28^\circ$, so probably
aligned with the planets and disk. Based on the assumption of star - planet - disk
alignment in `typical; (i.e. non hot Jupiter) systems, we suggest that the most probable
system-wide inclination can be inferred if the inclination of just one component has been
measured.

\section{The HD 82943 system}

\subsection{The Star}

HD 82943 is a nearby (27.5pc) Sun-like main-sequence dwarf star
(F9V). \citet{2004A&A...415..391M} quote an age of 2.9 Gyr, while
\citet{2009A&A...501..941H} derive an upper limit of 2.8 Gyr. The age is clearly
uncertain, but relatively unimportant for our analysis because it is only used in
considering how long the planets have had to influence the debris disk. We therefore
adopt an age of 3Gyr.

The stellar rotational velocity is $v \sin i = 1.35$-1.7 km s$^{-1}$
\citep{2004A&A...415..391M,2006ApJ...646..505B}. Using the inferred period of 18 days
\citep{2004A&A...415..391M} and the stellar radius of $1.15 R_\odot$ derived from SED
fitting (see section \ref{s:obs}), the inclination of the stellar pole from our line of
sight is $28 \pm 4^\circ$, if only the range of $v \sin i$ is used to calculate the
uncertainty. The rotation period was derived from the $R'_{\rm HK}$ activity indicator
rather than directly measured, which \citet{1984ApJ...279..763N} show results in period
uncertainties of a few days. A three day uncertainty yields an inclination uncertainty of
$\approx$5$^\circ$ here, so while direct verification of the period would be beneficial,
our derived inclination is unlikely to change significantly.


\subsection{The Planets}

Two $M \sin i \approx 1.8$ Jupiter-mass planets were discovered to orbit HD 82943 in 2004
\citep{2004A&A...415..391M}. The orbital periods are similar to the Earth's---219 and 435
days---meaning that these are not hot Jupiters. These planets were recognised to be in a
2:1 mean motion resonance, and studies followed that aimed to understand their dynamics
and the true constraints on the orbital parameters, even showing that the observed radial
velocities may be explained by two planets in a 1:1 resonance \citep[i.e. a Trojan
pair,][]{2005ApJ...621..473F,2006ApJ...641.1178L,2006ApJ...647..573G,2008MNRAS.385.2151B}. Where
they considered the 2:1 resonance, these studies did not consider the system inclination
relative to the sky plane. However, because they are in resonance and relatively massive,
the planets' mutual perturbations should result in significant departures from purely
independent Keplerian orbits. These departures are sensitive to the planet masses, hence
providing an opportunity to constrain the planet inclinations with sufficiently high S/N
data that spans sufficiently long time period \citep[e.g.][]{1992Natur.355..325R}.

Recently, \citet{2013arXiv1306.0687T} presented additional data for the HD 82943
system. Because more than eight orbital periods of the outer planet have now been
observed, they attempted to constrain the planetary inclinations. Their method involved
deriving rough orbital parameters using Keplerian orbits, and using these as a starting
point for a $\chi^2$ minimisation method using a dynamical model that accounts for
planet-planet interactions. With the assumption that the two planets are mutually aligned
(coplanar), they concluded that the most likely inclination of the two planets is near to
face-on, specifically at $20 \pm 4^\circ$. Naturally, the low inclination means that
$\sin i$ is relatively small, and that the planet masses are both quite hefty at 4.8
Jupiter masses. If the assumption of mutual alignment of the planets is relaxed,
\citet{2013arXiv1306.0687T} found that no useful inclination constraints could be made,
but they argued that mutual alignment is more plausible, essentially because the mutually
aligned model has fewer free parameters. The similar inclination measured for the debris
disk below adds strength to their conclusion of mutual planet alignment.

The inclination derived for the coplanar configuration is consistent with the stellar
inclination derived above. However, because neither the position angle of the stellar
pole nor the planetary line of nodes can be derived from the current observations, the
conclusion of alignment relies on the argument that it is unlikely that both inclinations
would be similar and close to face-on (there is a 0.5\% chance that two systems randomly
drawn from a distribution uniform in $\cos i$ will be between 20 and 30$^\circ$). To
independently derive the inclination of the planets would require either direct imaging
or astrometry, the latter being more likely given the small angular size of the planetary
orbits (though the perturbation is of order hundreds of micro-arcseconds).

\subsection{The Debris Disk}\label{s:obs}

\begin{figure}
  \begin{center}
    \hspace{-0.5cm} \includegraphics[width=0.5\textwidth]{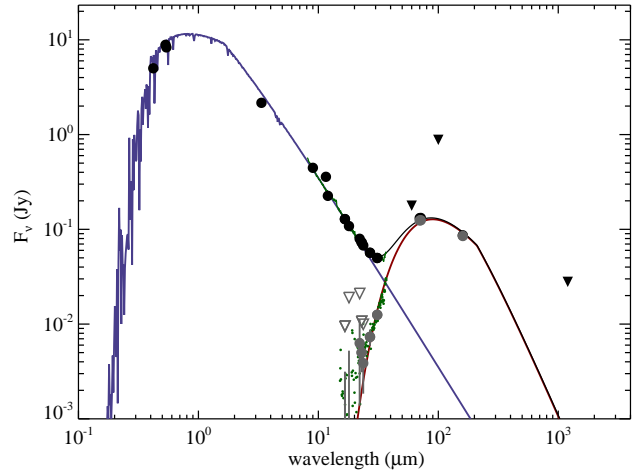}
    \caption{Spectral energy distribution for HD 82943. Dots are fluxes and triangles
      3$\sigma$ upper limits. Black symbols are measured fluxes and grey symbols are
      star-subtracted (i.e. disk) fluxes. The 5990K stellar photosphere model is shown in
      blue, the 57K blackbody disk model in red, and the star+disk spectrum in black. The
      green line shows the observed IRS spectrum, and the green dots show the
      star-subtracted spectrum.}\label{fig:sed}
  \end{center}
\end{figure}

\begin{figure*}
  \begin{center}
    \hspace{-0.25cm} \includegraphics[width=0.5\textwidth]{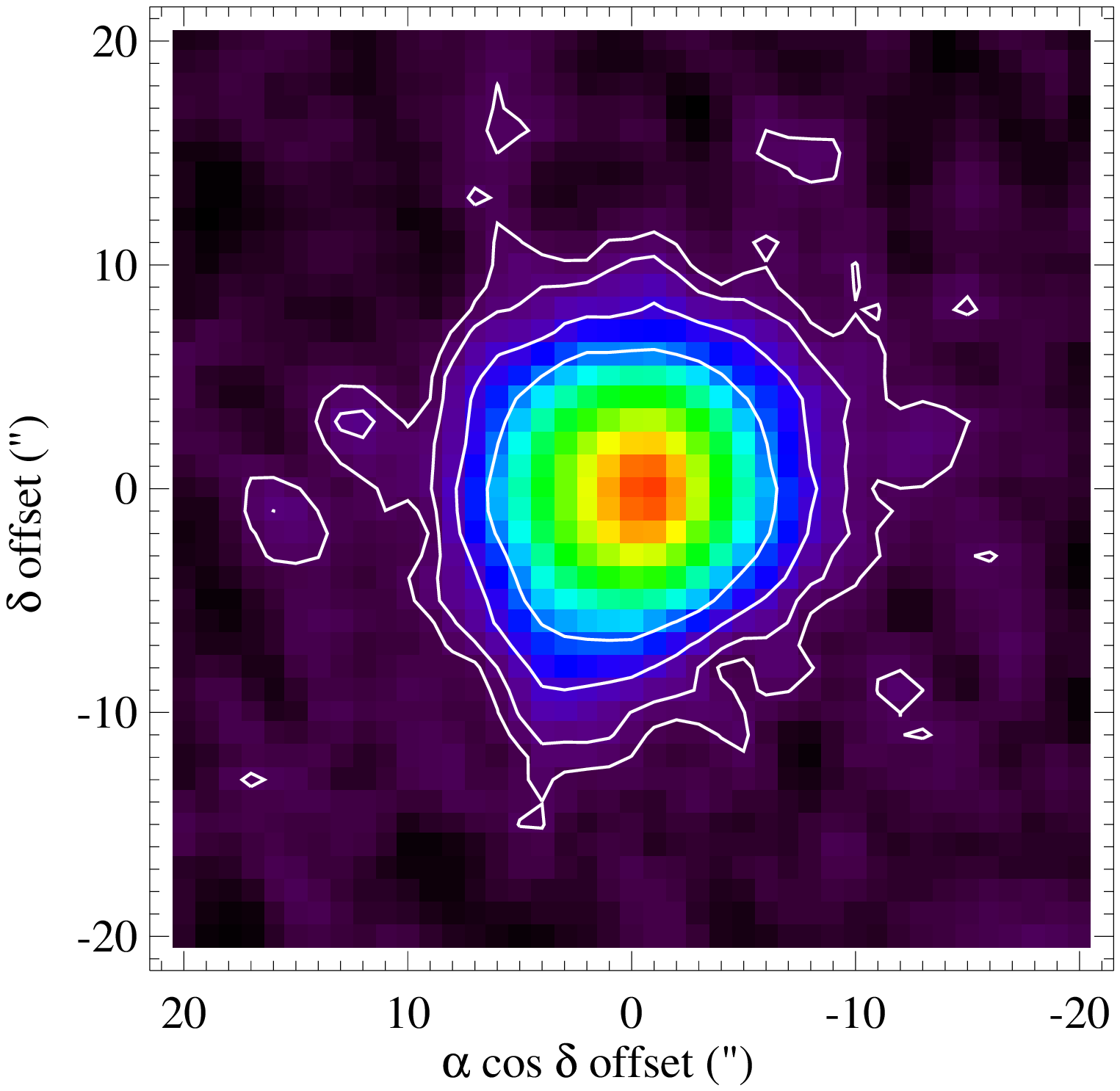}
    \hspace{-2cm} \includegraphics[width=0.5\textwidth]{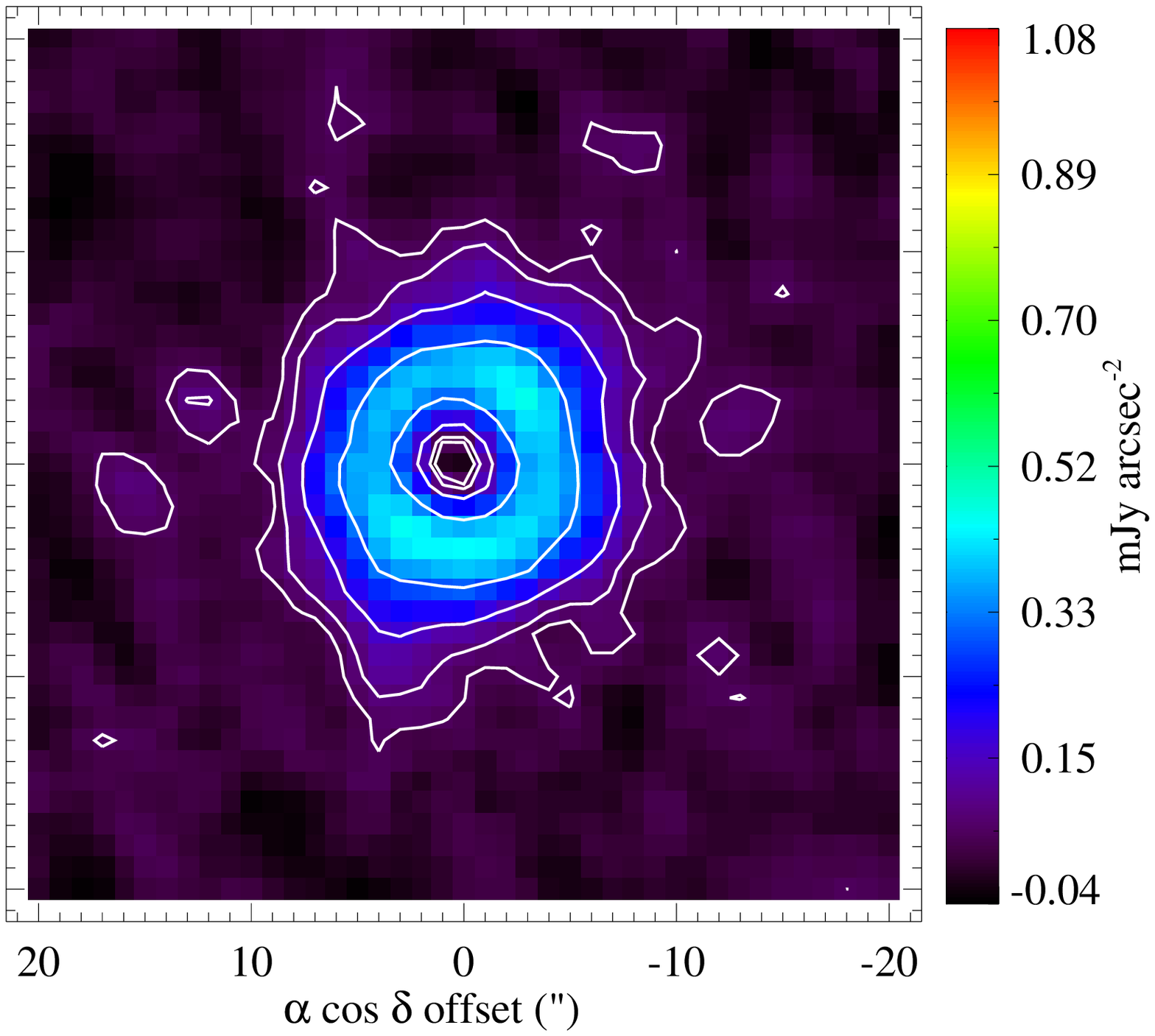}
    \caption{\emph{Herschel} 70$\mu$m images of HD 82943, North is up and East is
      left. The left panel shows the raw image with contours at 3, 5, 10, and 20 times
      the pixel RMS of $1.6 \times 10^{-2}$mJy arcsec$^{-2}$. The right panel shows the
      same image and contours after a peak-scaled point source has been subtracted,
      leaving near-circular residuals as a clear sign of a near face-on
      disk.}\label{fig:ims}
  \end{center}
\end{figure*}

The debris disk around HD 82943 was first discovered by \citet{2005ApJ...622.1160B}, as
part of a program to observe planet-host stars, with photometry using the Multiband
Imaging Photometer for \emph{Spitzer} \citep{2004ApJS..154....1W,2004ApJS..154...25R}. An
infrared excess above the stellar photosphere at 70$\mu$m was seen, with the excess
attributed to the presence of a significant surface area of small grains in a debris
disk. The excess was not detected at 24$\mu$m so the disk temperature and fractional
luminosity were not constrained (see their Fig 9). The system was subsequently observed
with the \emph{Spitzer} Infra-Red Spectrograph \citep[IRS,][]{2004ApJS..154...18H},
though the spectrum has never been published. Here, we use the CASSIS-processed version
of these data \citep{2011ApJS..196....8L}, which show a significant excess beyond about
25$\mu$m.

In November 2011, HD 82943 was observed by \emph{Herschel} \citep{2010A&A...518L...1P}
using the Photodetector and Array Camera \& Spectrometer (PACS) instrument \citep[][see
Table \ref{tab:obs}]{2010A&A...518L...2P} as part of the Search for Kuiper Belts around
Radial-velocity Planet Stars (SKARPS). The overall goal of the survey is to look for
correlations between debris disk and planet properties by observing systems known to host
planets discovered by radial velocity. The observations used the standard ``mini
scan-map'', which comprises two sets of parallel scan legs, each taken with a 40$^\circ$
difference in scan direction. The raw timelines were projected onto a grid of pixels
(i.e. turned into images) using a near-standard HIPE pipeline
\citep{2010ASPC..434..139O}. The fluxes at 70 and 160$\mu$m were measured using aperture
photometry (radii of 15'' \& 20''), yielding fluxes of $129 \pm 4$mJy and $87 \pm 7$mJy
at 70 and 160$\mu$m respectively.

Figure \ref{fig:sed} shows the spectral energy distribution for HD 82943, including the
\emph{Spitzer} and \emph{Herschel} data. We fit PHOENIX models from the Gaia grid
\citep{2005ESASP.576..565B} to optical and near-IR data using least squares minimisation,
finding a stellar effective temperature of 5990K and a radius of $1.15R_\odot$. We then
use the stellar photosphere model to predict the flux density at longer wavelengths
(e.g. $7.2 \pm 0.2$ and $1.35 \pm 0.03$mJy at 70 and 160$\mu$m), thereby demonstrating
that the \emph{Spitzer} and \emph{Herschel} data are significantly in excess of the level
expected. We fit a simple blackbody model to the excess fluxes, finding a fractional
luminosity of $L_{\rm disk}/L_\star=10^{-4}$ and a temperature of $57 \pm 2$K, with the
small uncertainty due to detection over a reasonably wide range of wavelengths
(20-160$\mu$m). In Figure \ref{fig:sed} we have multiplied the blackbody disk spectrum by
$( \lambda_0/\lambda)$ beyond $\lambda_0=210\mu$m \citep{2008ARA&A..46..339W}, to account
for inefficient long-wavelength emission by small grains and ensure a more realistic
prediction of the far-IR/sub-mm disk brightness. Assuming that it lies in a single narrow
ring, the blackbody temperature implies that the disk lies at a stellocentric radius of
30AU. We show below that the disk actually lies farther away, consistent with the bulk of
emission coming from grains that emit inefficiently at wavelengths longer than their
size, which must emit at hotter-than-blackbody temperatures to maintain energy
equilibrium.

\begin{table}
  \caption{\emph{Herschel} observations of HD 82943. Each Obs ID represents a single
    scan direction, and the two differ by 40$^\circ$.}\label{tab:obs}
  \begin{tabular}{llll}
    \hline
    ObsID & Date & Instrument & Duration (s) \\
    \hline
     1342232212 & 10 Nov 2011 & PACS 100/160 & 1686 \\
     1342232213 & 10 Nov 2011 & PACS 100/160 & 1686 \\
    \hline
  \end{tabular}  
\end{table}

In addition to yielding photometric measurements, the disk is well resolved by
\emph{Herschel} at 70$\mu$m, but less so at 160$\mu$m. There is in addition some apparent
low-level background contamination to the NE at 160$\mu$m. Such contamination is in fact
fairly common for \emph{Herschel} observations at this wavelength; here we are less than
a factor of two above the confusion limit of 1.4mJy (as predicted by the \emph{Herschel}
Observation Planning Tool). The 70$\mu$m image is shown in the left panel of Figure
\ref{fig:ims}. To show that the image is resolved, the right panel shows the image after
a peak-normalised point source (calibration star $\gamma$ Dra, processed in the same way
as the data and rotated to the same position angle) was subtracted, leaving a clear ring
of extended emission. In addition to showing that the disk is resolved, the azimuthal
symmetry of the remaining ring shows that the disk is near to face-on.

To estimate the inclination and position angle of the disk we use two independent
methods. The first is simple, we fitted a 2D Gaussian to the star-subtracted image of the
HD 82943 disk, finding a position angle of 147$^\circ$ and an inclination of
30$^\circ$. The inclination is found using $\cos i = s_{\rm min}/s_{\rm maj}$, where
$s_{\rm maj}$ and $s_{\rm min}$ are found from quadratically subtracting the PACS
70$\mu$m beam full-width at half-maximum (FWHM) of 5\farcs75 from the major and minor
components of the fitted Gaussian FWHM ($s_{\rm maj}$ is also an estimate of the
characteristic disk size, about 100AU). To estimate the uncertainty we then added the
Gaussian fit image into an off-center position in 9 other 70$\mu$m observations from our
programme (all observations have the same depth). A Gaussian was then fitted at this
position and the position angle and inclination derived. This method is a simple way of
estimating how the disk geometry can vary due to different realisations of the same noise
level. The inclinations vary from 25-31$^\circ$ with a mean of 28$^\circ$, while the
position angles vary from 133 to 153$^\circ$ with a mean of 147$^\circ$.

As a second method we fit a physical model for the disk structure and estimate parameter
uncertainties in a more traditional way. These models have been used previously to model
\emph{Herschel}-resolved debris disks
\citep[e.g.][]{2012MNRAS.426.2115K,2013ApJ...762...52B}, and generate a high resolution
image of an azimuthally symmetric dust distribution with a small opening angle, as viewed
from a specific direction. These models are then convolved with a PSF model for
comparison with the observed disk. The best fitting model is found by a combination of
by-eye coaxing and least-squares minimisation. We found that the HD 82943 disk could not
be well modelled by a simple ring, and hence use a dust distribution that extends from 67
to 300AU, with the face-on optical depth distributed as a power-law that decays as
$r^{-1.6}$ and is normalised to be $3.98 \times 10^{-4}$ at 1AU. The temperature
distribution is assumed to decay as $r^{-0.5}$ (i.e. like a blackbody, which is 278.3K at
1AU), but is required to be hotter at the same distance by a factor $f_{\rm T}=1.8$
(i.e. 567K at 1AU) to reconcile the temperature of the SED with the observed radial
location of the dust \citep[see][]{2012MNRAS.424.1206W,2012A&A...548A..86L}. That this
factor is larger then unity is consistent with the result that the inner disk radius is
significantly larger than the radius implied by the simple blackbody SED model, because
it is also a signature of inefficient long-wavelength grain emission and small grains
dominating the disk emission. The best disk model is inclined by 27$^\circ$ at a position
angle of 152$^\circ$, and the residuals when the best fitting model are subtracted from
the data show no significant departures from the background noise elsewhere in the map.

\begin{figure}
  \begin{center}
    \hspace{-0.5cm} \includegraphics[width=0.5\textwidth]{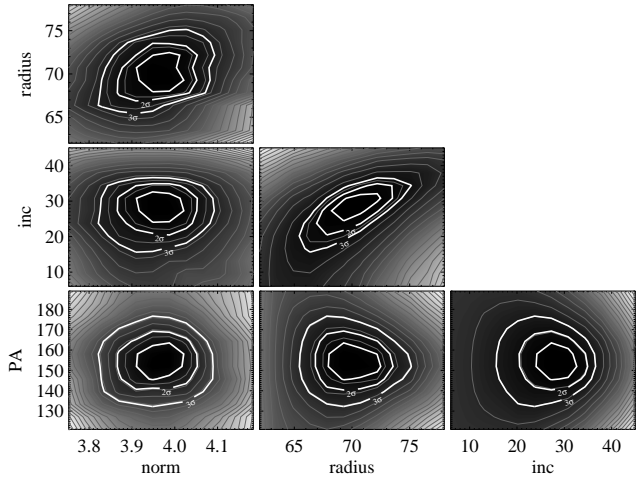}
    \caption{$\chi^2$ contours for varying disk normalisation ($\times 10^4$), inner
      radius (in AU), inclination, and PA (both in degrees). Each panel shows contours
      for two parameters when marginalised over the other two.}\label{fig:cont}
  \end{center}
\end{figure}

To estimate the uncertainty in several parameters, we then calculate a grid around the
best fit location, varying the disk normalisation, the inner radius, the inclination, and
the position angle. Each parameter is calculated at 12 values, giving a grid with 20,736
models. For each model we calculate the $\chi^2$ from the model-subtracted residuals,
accounting for correlated noise by increasing the noise by a factor of 3.6 over the
pixel-to-pixel RMS \citep[see][]{2002PASP..114..144F,2012MNRAS.421.2264K}. The results of
this grid calculation are shown in Figure \ref{fig:cont}, where the white contours show
$\Delta \chi^2$ values corresponding to 1, 2, and 3$\sigma$ departures from the best
fit. The inclination is constrained to $27 \pm 4^\circ$, while the PA is $152 \pm
8^\circ$. These estimates agree well with the simple Gaussian fitting, with the
difference in the range of position angles most likely because the PACS beam is slightly
elongated, which will influence the results from naive Gaussian fitting. While the
position angle is not particularly well constrained, we conclude that the inclination is.

The disk inclination is therefore similar to that of both the star and the planets. While
the line of nodes has only been derived for the disk, we take the similar inclinations to
be highly suggestive of system-wide alignment. The chance of three randomly drawn
inclinations to all be between 20 and 30$^\circ$ is 0.04\%, so the star - planet - disk
alignment is very unlikely to be coincidental.

Combined with the possible near face-on planet orbits, one question is then whether the
likely planet - disk alignment is due to nature or nurture. Given the adopted system age
of 3Gyr and the relatively massive planets, it may be that secular perturbations have
over time pulled the average inclinations of parent bodies in the debris disk into
alignment from an initially misaligned configuration. If this were the case, then the
alignment of the planets and disk would be required by the dynamics if no other forces
are acting. If the disk is too distant to have been affected, the alignment can be
considered primordial and be used as evidence that disk-planet alignment was the natural
outcome in this system.

\begin{figure}
  \begin{center}
    \hspace{-0.5cm} \includegraphics[width=0.5\textwidth]{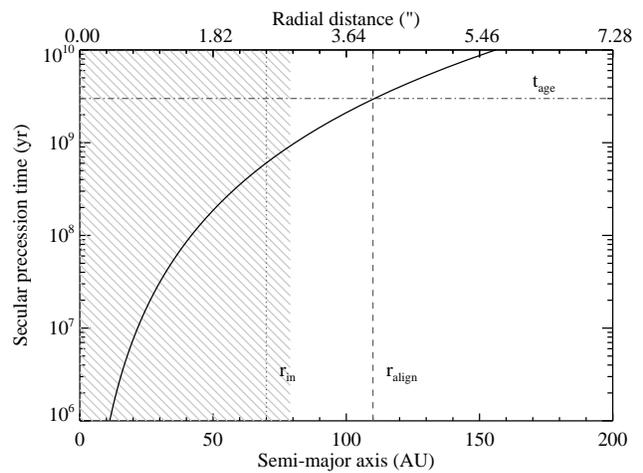}
    \caption{Secular precession time for planetesimals due to the outer planet (assuming
      4.8 Jupiter masses). Planetesimals can start to be aligned (i.e. have executed one
      cycle of secular precession) within 3Gyr if they reside within 110AU. Beyond 110AU,
      where the bulk of the resolved disk emission lies, the disk is not significantly
      affected and thus the alignment is primordial.}\label{fig:tsec}
  \end{center}
\end{figure}

A comparison of the secular precession time due to the outer planet with the system age
and disk size is shown in Figure \ref{fig:tsec}. The secular precession time is
calculated according to \citet{2010MNRAS.401.1189F}, and the black line shows the radius
at which particles will undergo one precession period as a function of age. The hashed
area shows where the disk is within one half-width half-maximum of the \emph{Herschel}
PACS 70$\mu$m beam, and hence approximately where the disk inclination is
unconstrained. The disk inner edge at 67AU is marked, as is the radius of 110AU at which
disk particles have undergone one secular precession cycle at the stellar age of 3Gyr
(called $r_{\rm align}$). The disk outer radius is poorly constrained because the
power-law decay of the optical depth fades with increasing distance, but Figure
\ref{fig:ims} shows that significant surface brightness exists out to at least 10''
(275AU) in radius, well beyond the maximum distance where the disk could be aligned by
secular perturbations. Though the stellar age is also uncertain, this uncertainty is
unlikely to be important. For significantly younger ages the disk would be aligned to
smaller distances than 110AU. Even for an age of 10Gyr the disk would only be affected
out to 150AU. To check that the inclination derived for the outer disk is not simply
influenced by higher S/N in the inner regions, we created a model in which the disk was
separated at 110AU into two radial components, with each having independent
inclinations. In the best fitting model after $\chi^2$ minimisation the difference in
inclinations for the two components is less than 1$^\circ$. Thus, we conclude that the
inclination of the disk is independent of the inclination of the known planets, and
therefore that any planet - disk alignment is primordial.

\section{Conclusions}\label{s:disc}

We have shown that the debris disk surrounding HD 82943 is near face-on, with an
inclination of $27 \pm 4^\circ$. Assuming that the planet orbits are coplanar, the likely
planet orbit inclinations of $20 \pm 4^\circ$ and the inferred stellar inclination of
$28^\circ$ argue for primordial system-wide alignment at a level similar to the Solar
System. Though the line of nodes can only be derived for the debris disk, the chance of
all three components randomly having near face-on inclinations is about 0.04\%.

As a rough estimate of the number of other planetary systems in which long-term radial
velocity monitoring might be used to derive system inclinations, 33/90 systems with two
or more planets in the Exoplanet Orbit Database\footnote{On 11 April 2013}
\citep{2011PASP..123..412W} have maximum/minimum period ratios less than 2.3. While the
perturbations in many of these systems will may not be detectable, at least some should
allow inclination measurements similar to that made for HD 82943.

There are of course other possibilities for testing system alignment, with perhaps the
best tests being in edge-on systems. For example, an edge-on disk is the best place to
look for out-of-plane perturbations, such as the warp seen in the $\beta$ Pictoris
disk. These systems are also needed to use the Rossiter-McLaughlin effect to test for
star - planet misalignment.

In the absence of evidence for strong dynamical influences, such as those that may form
hot Jupiters, it seems that a picture of general alignment is emerging in extra-Solar
planetary systems. However, given that the first hot Jupiters were also found to be
aligned more systems need to be tested. If the trend of alignment continues, it will
argue strongly that measurement of the inclination of any component of the planetary
system, including the star itself, can act as a proxy for the inclination of the system
as a whole.

\section{Acknowledgements}

We thank the referee for a concise review and help on transiting planet details. This
work was supported by the European Union through ERC grant number 279973 (GMK \&
MCW). This research has made use of the Exoplanet Orbit Database and the Exoplanet Data
Explorer at exoplanets.org.


\end{document}